\documentclass[debug]{rmaa}

\usepackage{paralist}

\usepackage{psfrag,color}

\usepackage[latin1]{inputenc}
\title{New Spectral Analysis Results Within the Scope of Extended Matter Research in the AR Lacertae Active Binary System} 

\author{
  O. Karaku\c{s},\altaffilmark{1} 
  and F. Ekmek\c{c}i,\altaffilmark{1}
  }

\altaffiltext{1}{Ankara University, Faculty of Science, Department of Astronomy and Space Sciences, 06100 Tando\u{g}an, Ankara, Turkey}

\shortauthor{Karaku\c{s} \& Ekmek\c{c}i}
\shorttitle{New Spectral Analysis Results of AR Lac}

\fulladdresses{
\item Osman Karaku\c{s}: Ankara University, Faculty of Science, Department of Astronomy and Space Sciences, 06100 Tando\u{g}an, Ankara, Turkey (o\_karakus@yahoo.com).
\item Fehmi Ekmek\c{c}i: Ankara University, Faculty of Science, Department of Astronomy and Space Sciences, 06100 Tando\u{g}an, Ankara, Turkey \\(fekmekci@science.ankara.edu.tr).}

\listofauthors{W. J. Henney, A. Collaborator, \& L. Author}
\indexauthor{Henney, W. J.}
\indexauthor{Collaborator, A.}
\indexauthor{Author, L.}

\abstract{Within the scope of extended matter research, we present new spectral analysis results of an active binary system AR Lac. The low and high resolution spectra of this system  were taken during the period 2013-2016. The evaluation of low dispersion spectra together with the \texttt{B, V, R$_{c}$, I$_{c}$} and \texttt{WISE} photometric data showed that AR Lac has an excess radiation in \texttt{W2} band. In addition, the spectral energy distribution and the minima depth ratios of the light curves of this active binary system were studied to examine the flux contributions of the components of the system depending on wavelengths and on orbital phase. Furthermore, high resolution spectral analysis showed evidence of prominence-like structures and a possible extended matter around the cooler component of AR Lac binary system.}

\resumen{......}

\addkeyword{Stars: activity}
\addkeyword{(Stars:) binaries: eclipsing}
\addkeyword{Stars: chromospheres}
\addkeyword{(Stars:) circumstellar matter}
\addkeyword{Stars: individual: AR Lac}

\begin{document}
% Typeset article header
\maketitle

\section{Introduction}
\label{sec:intro}

AR Lac (HD 210334, Vmax = 6.09, P = 1.98 days) is one of the bright and totally eclipsing binary system of chromospherically active binaries \texttt{(CABs)}. This RS CVn type binary system has components of spectral type \texttt{G2 IV + K0 IV}. \citet{Luetal2012} summarized the previous studies that had been presented by numerous investigators, including photometric and spectroscopic observations of AR Lac. This active binary system is well known to have orbital period change, magnetic and spot activity which are affecting the cooler component together with strong emission features in Ca II H and K  lines \citep[see][]{Luetal2012}. A long-term secular period decrease was estimated at a rate of $\mathrm{dP/dt} = -(2.128\pm0.060)\times10^{-9} d/d$ by \citet{Luetal2012}, which may be caused by the magnetic activity of this active binary system. Based on this result, they also gave an estimation on the mass-loss rate for this binary system as $\mathrm{dM/dt} = -2.8\times10^{-10} M_\mathrm{\odot} yr^{-1}$. 

The spectral H$\alpha$ and Ca II H and K emission lines are very important indicators of chromospheric or magnetic activity of the component(s) of \textit{CABs} \citep [see][]{Rodono1980}. Thus, the activity level of a star can also be inferred by determining the presence of spectral H$\alpha$ emission line or by determination of H$\alpha$ with a core which was filled-in some absorption features \citep[see][]{Fernandezetal1994, Barden1985, Fekeletal1986, Boppetal1988, Strasetal1990}. 

In the studies of some \texttt{CABs}, it was found that there were some evidences which indicate that the extended/circumstellar matter in that binary system may exist \citep[see][]{Scaltritietal1993}. In these studies the measurements of the excess radiation, especially in the spectral region of long wavelengths of active binary systems, were used with the respect to excess radiation, may be caused by mass loss due to stellar winds, and which may related to  extended/circumstellar matter. 

Assessments of some important observational data of disk structure around chromospherically  active binary stars began with the infrared astronomical satellite \texttt{(IRAS)} in 1983. \texttt{IRAS} was the first mission to put a telescope in space to survey the All Sky Survey at 12, 25, 60, and 100 micron bands \citep{Scaltritietal1993}. Based on \texttt{IRAS} observations, \citet{Bussoetal1988} found that \texttt{IR} excess is definitely present in CF Tuc, while the spectral distributions of the $\lambda$ And, UX Ari and AR Lac can be accounted for \texttt{IR} excess by combinations of normal stellar components. They also concluded that the excess is not correlated with the activity level nor with the evolutionary status but may be correlated with the mass loss phenomena near the Main Sequence. Possible interpretation of the excess emission based on the evolutionary status of the binary components are also discussed by \citet{Bussoetal1990}. They discuss possible explanations in terms of mass loss phenomena(triggered by the binary nature) during the evolution of the sources near the Main Sequence. The behaviour of excess \ion{Ca}{ii} H and K and H$\epsilon$ emission in a sample of 73 \texttt{(CABs)}, including AR Lac, was examined by \citet{Montesetal1996}, and they found that there was a good correlation between excess \ion{Ca}{ii} K and H$\epsilon$ chromospheric emission fluxes.

In the H$\alpha$ line study of the system \citet{Frascaetal2000} found that there are chromospheric emissions from both components in most spectra and the rotational modulation of H$\alpha$ line emission is not obvious. They gave an interpretation on the excess absorption observed in 1997 during and near the primary minima as a comment for the effect of a prominence like structure anchored between the leading and trailing hemisphere of the cool component. This interpretation was fully compatible with the radial velocities of H$\alpha$ peaks in extracted spectra. \citet{Zboriletal2004} reported that the central depth of H$\alpha$ profiles of AR Lac at eclipses (at 0.041P and 0.043P) were deeper than synthetic profiles and close to the profiles of stars with the same spectral types(e.g. $\delta$ Eri).

\citet{Lanzaetal1998} gave a detailed analysis of the long-term and seasonal light curves of AR Lac. Based on their main results concerning the magnetic activity of AR Lac binary system, they concluded that the large active region around the substellar point on the secondary showed itself not only at photospheric levels but also in the chromosphere and corona, with an extended structure which might well be an interconnecting loop between the two stars. 

The results of \texttt{Very Large Array (VLA)} observations during optical eclipses in 1977 of AR Lac were reported by \citet{OandS1977}. They found a small increase in radio flux density during 0.5P, although the radio source had variation on time scales $\geq$ a few hours, but they did not detect a strong eclipse-like feature that occurs near 0.0P or 0.5P orbital phases. Together with their spectral evaluations by taking into account the synchrotron self-absorption, they suggested that the radiation is likely to be produced in a volume much larger than the stars in the system due to the lack of clearly defined eclipse in the AR Lac binary system.

Simultaneous observations of AR Lac at radio and ultraviolet wavelengths during two consecutive secondary eclipses, made in 1979 May, were reported by \citet{Brownetal1979}. During one of the eclipses they saw a quiescent radio source without an evidence of an eclipse at radio wavelengths. On the second one, they found the radio source was brighter by a factor of 3 together with an evident radio eclipse. And, in the latter case, the radio source was circularly polarized by 5-10\% during eclipse. They also gave the following with the same statements as follows:
\begin{itemize}
\item[-]
the sense of circular polarization change abruptly at first contact, and
\item[-]
the radio eclipse was twice as deep in one sense of circular polarization as it was in the other.
\end{itemize}
They also identified the following features from their simultaneous ultraviolet observations:
The ultraviolet coronal lines also showed the eclipse. But, the eclipse on the day in which the radio source was quiescent became much more prominent than the eclipse on the day in which the radio source was active. In other words, the effect of eclipse on coronal ultraviolet lines increases while the radio source is in quiescent state, but while the radio source is in an active state the effect of eclipse decreases and the eclipse becomes more uncertain. Thus, this result is evident that coronal activity in the radio region increases the brightness and this increase will not only be caused by stellar activity because the brightness effect of the radiation in ultraviolet region has not been seen at the same time.

\begin{table}
\footnotesize
\caption{Log of \texttt{TUG TFOSC} spectral observations of three standard stars and AR Lac active binary system.}
 \label{tab1}
 %\begin{threeparttable}[t]
 \centering
 \setlength{\tabcolsep}{0.5\tabcolsep}
 \begin{tabular}{lllllcccc}
 \hline
 Stellar & Type & Date & HJD & Start time & Exposure time & Airmass & Orbital & Number \\
Object & & of obs. & & of obs.(UT) & (sec.) & (mag.) & phase & of images \\
\hline
Vega & \textit{Standard} & Oct. 12, 2015 & 2457308.278	& 18:40:02 & 0.8 &	1.316 &	- &	5 \\
AR Lac & \textit{Variable} & Oct. 12, 2015 & 2457308.428 & 22:11:28	& 30 & 1.295 &	0.995 & 10 \\
HR 5510 & \textit{Reference} & June 04, 2016 & 2457544.282	& 18:42:21 & 5 & 1.033 &	- &	5 \\
Vega & \textit{Standard} & June 04, 2016 & 2457544.351	& 20:21:48 & 5 &	1.286 &	- &	5 \\
Vega & \textit{Standard} & June 05, 2016 & 2457545.404	& 21:39:36 & 0.03 &	1.089 &	- &	10 \\
AR Lac & \textit{Variable} & June 05, 2016 & 2457545.422 & 22:08:49 & 20 & 1.739 &	0.496 & 10 \\
HR 8634 & \textit{Standard} & Sept. 05, 2016 & 2457637.322	& 19:35:20 & 2 & 1.271 &	- &	10 \\
AR Lac & \textit{Variable} & Sept. 05, 2016 & 2457637.438 & 22:25:47 & 7 & 1.048 &	0.894 & 10 \\
\hline
    \end{tabular}
\end{table}

\begin{table}
\footnotesize
\caption{Log of \texttt{TUG} Echelle spectral observations of AR Lac active binary system.}
\label{tab2}
 %\centering
 \setlength{\tabnotewidth}{1.1\linewidth}
 \setlength{\tabcolsep}{0.5\tabcolsep} \tablecols{9}
 \begin{tabular}{lllcccccc}
 \hline
  & & Start time & Exposure & Airmass & & \texttt{G2 IV}& \texttt{K0 IV} &	\\
HJD & Date of obs. & of obs. & time & (mag.) &	Phase\tabnotemark{a} & Contribution & Contribution & S/N \\
 & & (UT) & (sec.) & & & rate & rate & \\
\hline
2456487.407	&	13.07.2013	&	21:29:29	&	1800	&	1.242	&	0.012	& 0.00	&	1.00	&	41	\\
2456487.437	&	13.07.2013	&	22:04:36	&	2700	&	1.157	&	0.026	& 0.08	&	0.92	&	54	\\
2456487.475	&	13.07.2013	&	22:59:53	&	2700	&	1.069	&	0.046	& 0.25	&	0.75	&	84	\\
2456487.509	&	13.07.2013	&	23:48:50	&	2700	&	1.027	&	0.063	& 0.34	&	0.66	&	83	\\
2456487.542	&	14.07.2013	&	00:36:32	&	2700	&	1.012	&	0.080	& 0.36	&	0.64	&	92	\\
2456487.576	&	14.07.2013	&	01:24:42	&	2700	&	1.023	&	0.097	& 0.36	&	0.64	&	94	\\
2456488.401	&	14.07.2013	&	21:12:38	&	2700	&	1.279	&	0.513	& 0.46	&	0.54	&	90	\\
2456488.434	&	14.07.2013	&	22:00:09	&	2700	&	1.158	&	0.529	& 0.45	&	0.55	&	90	\\
2456488.467	&	14.07.2013	&	23:01:02	&	2700	&	1.040	&	0.546	& 0.44	&	0.56	&	94	\\
2456488.501	&	14.07.2013	&	23:36:17	&	2700	&	1.032	&	0.563	& 0.37	&	0.63	&	70	\\
2456488.534	&	15.07.2013	&	00:23:47	&	2700	&	1.013	&	0.579	& 0.36	&	0.64	&	94	\\
2456488.567	&	15.07.2013	&	01:11:18	&	2700	&	1.019	&	0.596	& 0.36	&	0.64	&	80	\\
\hline
\tabnotetext{a}{Orbital phases are computed by using the light elements taken from TIDAK (TIming DAtabase in Krakow) http://www.as.up.krakow.pl/ephem/old-ephem/EPHEM-2012xi.TXT.}
 \end{tabular}
 \end{table}

By using the \texttt{Einstein} Imaging Proportional Counter \texttt{(IPC)} observations of the X-ray eclipse together with the cotemporal \texttt{International Ultraviolet Explorer (IUE)} and radio observations, \citet{Walteretal1983} found an extended corona with a scale of about 1R$_{*}$ to be associated with the \texttt{K0 IV} component of AR Lac binary system. They have obtained X-ray, radio, and ultraviolet observations of the system over one orbital cycle. The primary minimum of the X-ray light curve they obtained for AR Lac was deep and the secondary minimum was broad and shallow. They found that quiescent corona of the \texttt{G2 IV} component is small and asymmetric which extends to some 0.02R$_{*}$ above the photosphere to be related with stellar spots. On the other hand, the \texttt{K0 IV} component was found to have two coronal components: 1- An outer, extended coronal component which is presumably a hotter component that extends to 1R$_{*}$ above the photosphere and exhibits a bright hemisphere, 2- An inner coronal component which is small relative to the stellar radius.

Using their \texttt{VLA} observations at 1.5 and 4.9 GHz frequencies on 13 and 15 October 1982, \citet{DandM1984} have not detected a clear eclipse signature in the light curve of AR Lac. On both observing days, a significant circular polarization of 2\%-8\% was observed with a helicity reversal between 1.5 and 4.9 GHz, and they attributed this result to a gyro-synchrotron mechanism.

Using the \texttt{IUE LWR} and \texttt{SWP} spectra of AR Lac, obtained on 3-5 October 1983, and their \texttt{VLA} observation on 4-5 October 1983 at 2, 6, and 20 cm, \citet{Walteretal1987} determined the atmospheric structure within the plage regions together with the properties of the extended coronal component around \texttt{K0 IV} star. They identified three discrete regions of emissions in the outer atmosphere of the \texttt{K0 IV} star in which there are two "plages" and a chromospheric brightening that was related to a radio flare.

Based on their results of the multifrequency \texttt{VLA} and \texttt{Very Long Baseline Array (VLBA)} observations, made in 1997, \citet{Trigilioetal2001} inferred the conclusions as follows:
\begin{itemize}
\item["-]
the spectral and spatial information of the corona of AR Lac indicate a structured morphology, which can be modeled with a core-halo source;
\item[-]
the physical parameters, as derived from the fit of the observed spectra with the core-halo model, are consistent with the hypothesis of a co-spatial X-ray and radio source;
\item[-]
the observed radio emission cannot be attributed to the same thermal electron population responsible for the observed X-ray emission."
\end{itemize}

\citet{Koch2007} detected a variable polarization in the AR Lac. He reported that the seat of the polarization could provisionally be located in the assorted active clouds that populate the outer envelope of \texttt{K0 IV} component. Also, he reported that the plasma must be low in metals compared to the \textit{Sun} and not have a simple polarization spectrum. In this regard, he also gave the following explanation: "Since Z is about 0.6Z$_{\odot}$, there is a possibility that the metal depletion is only apparent and not real because the grain condensation in the cool envelope and that grains alone are the seats of the polarization."

Recently, a research on the extended/circumstellar matter in \texttt{CABs} has been presented by \citet{KandE2020}. Based on their photometric \texttt{CCD} data of 13 \texttt{CABs}, including AR Lac and together with the \texttt{2MASS} and \texttt{WISE} data, they found that the AR Lac had some fluctuations in colour excess \texttt{(CE)} values at around primary minimum. And, they also found that the system showed some characteristic variations in \texttt{CE} values towards longer wavelengths, at both around secondary minimum and outside eclipse but with lower and some detectable \texttt{CE} values in all bands at outside eclipses. They concluded that the main source of excess radiation in the AR Lac is stellar activity. This stellar activity is expected to contribute to the formation of an extended/circumstellar matter in the AR Lac, but a significant evidence on how much it contributes could not be presented by \citet{KandE2020}, based solely on their results of photometric \texttt{CE} measurements.

% ----- Table 3 -----

\begin{table}
\footnotesize
  %\centering
  \setlength{\tabcolsep}{0.5\tabcolsep} \tablecols{7} 
  \caption{Log of TUG Echelle spectral observations of reference stars.}
  \label{tab3}
  \begin{tabular}{lllllll}
  \hline 
 Star &	HJD & Date of obs. & Start time of obs.(UT) & Exposure time (sn)&	Airmass (mag.) & S/N \\ 
\hline
HR 6256	&	2456488.287	&	14.07.2013	&	18:36:28	&	1800	&	1.016	&	89	\\
HR 6256	&	2456488.310	&	14.07.2013	&	19:08:57	&	1800	&	1.007	&	61	\\
HD 195405	&	2456487.333	&	13.07.2013	&	19:39:31	&	2000	&	1.280	&	30	\\
HD 195405	&	2456487.366	&	13.07.2013	&	20:18:08	&	3000	&	1.174	&	31	\\
\hline
  \end{tabular}
   \end{table}

In this study, the spectra of AR Lac taken during the observing period 2013-2016 at different orbital phases by using the Coud\'{e} Echelle and \texttt{TFOSC (Faint Object Spectrograph and Camera)} instruments were analysed to investigate and to reveal the relationship between its stellar activity and a possible extended/circumstellar matter of the system. In addition, together with the results of this spectral analysis of the AR Lac system, an evaluation was made by including the results given by \citet{KandE2020} based on their photometric \texttt{CCD, 2MASS} and \texttt{WISE} data taken during minima and outside eclipses. Within the scope of this evaluation, the \texttt{B, V, R$_{c}$, I$_{c}$}, and \texttt{W1, W2, W3, W4 of WISE} band light curves of AR Lac were also taken into consideration. 

\section{Observations and Data Reductions}
\label{sec:obsdatred}

Spectral observations of AR Lac were carried out at times to cover the external and internal contact parts of the light curve of the system in the eclipses together with descending and ascending parts. The choice of these observation times corresponding to these orbital phases was made to see the relationship of the system's activity events with the characteristics of the component stars on the disc edges and a possible extended/circumstellar matter. For this purpose, both low and high resolution spectral data were used to investigate the presence of a possible extended/circumstellar matter together with the properties and evolutionary states of components of AR Lac.

\begin{figure}
\vspace{0.0cm}
\begin{center}
  
     \includegraphics[width=0.75\textwidth]{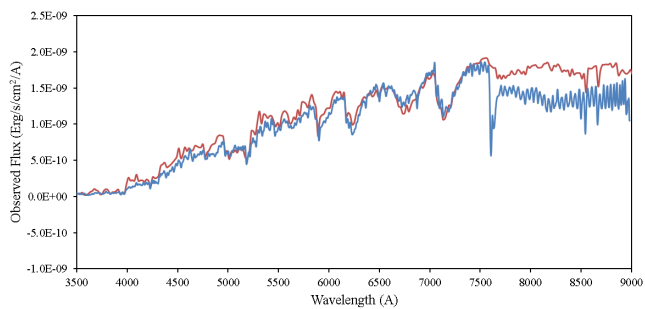}\\
     
\caption{A comparison synthetic spectrum(as red line) and observational \texttt{TFOSC} spectrum(as blue line) of HR 5510 reference stars. Dropping in the continuum fluxes due to telluric lines are seen clearly. Synthetic spectrum is constructed by taking $\mathrm{log} \mathit{g}=1.5$, $T(K)=3500$, and $Z=0$.}
\label{fig1}
\end{center}
\end{figure}

For all low and high resolution spectra of this study, Image Reduction and Analysis Facility \texttt{(IRAF)}\footnote{http://iraf.noao.edu} was used to reduce the combined data with standard procedures, including the corrections for flat and bias, and determination of aperture, wavelength calibration, velocity correction, and interstellar extinction/cosmic radiation. But, the atmospheric and interstellar extinction effects have only been corrected for low resolution spectra. The standard stars selected and observed within the conditions of observability for flux calibration are as follows:
Vega (HR 7001, \texttt{A0 V}, V = 0.03) and HR 8634 (\texttt{B8 V}, V = 3.41). Standard flux values for the stars are taken from the website "https://snfactory.lbl.gov/snf/spstds/" and the references therein. In addition, the \texttt{TFOSC} spectra of the HR 5510 (\texttt{M1 III}, V = 6.28) reference star were taken to see and evaluate the effect of Telluric line absorptions on the spectra of the AR Lac binary system. For H$\alpha$ profile analysis, the high resolution spectra of two reference stars which are of the same spectral type of both components of the AR Lac binary system, were also taken. These two stars are: HD 195405 \texttt{(G2 IV)} and HR 6256 \texttt{(K0 IV)}.

In addition, photometric data, from \citet{KandE2020}, of AR Lac in Johnson-Cousins \texttt{BVR$_{c}$I$_{c}$} and medium \texttt{WISE} bands were used to make a comparison with our spectroscopic results. In all these bands, the photometric light curves of the AR Lac were also constructed and evaluated by using the photometric colour excess measurements of \citet{KandE2020}. Some details about these photometric observational data are presented below in Section~\ref{sec:phwo}.

\begin{figure}
\vspace{0.0cm}
\begin{center}
  
     \includegraphics[width=0.75\textwidth]{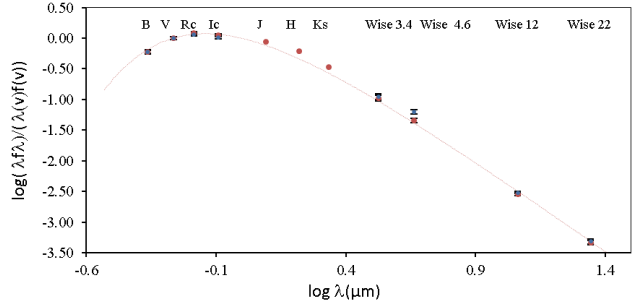}

\caption{Comparison of \texttt{SEDs} of AR Lac and reference star HD 56168. The photometric values of HD 56168 ($(B-V)_{\circ}$=$0.90\pm0.01$) are indicated as red points, while the values of AR Lac ($(B-V)_{\circ}$=$0.87\pm0.04$), during 0.0P, are indicated as blue points. The straight line shows the black body energy distribution of  \texttt{T(K)=5100}.}
\label{fig2}
\end{center}
\end{figure}

\subsection{Low Dispersion Spectra}
\label{sec:lds}

The optical low resolution spectral observations of AR Lac, a reference star(HR5510) and two standard stars(Vega and HR 8634) took place during the observing period 2015-2016, by using the \texttt{TUG} (T\"{U}B\.{I}TAK National Observatory) Faint Object Spectrograph and Camera \texttt{(TFOSC)} mounted on the 1.5 m Russian Turkish Telescope \texttt{RTT150}\footnote{Specifications of \texttt{RTT150} and \texttt{TFSOC} are available at www.tug.tubitak.gov.tr} in Antalya, Turkey. The log of \texttt{TFOSC} observations are given in Table~\ref{tab1}. Grism 15 was used with a 100 micron slit. The wavelength range in this configuration is $3230-9120 \AA$, and the resolving power (R)$\sim$749. Ar, Ne, He and Halogen lamp spectra taken in the same night with the star were used for wavelength calibration and flat-fielding. A total of 65 slit spectra were obtained: 30 of spectra were taken for AR Lac, and the rest were spectra of other standard stars (see Table~\ref{tab1}).

\begin{figure}
\vspace{0.0cm}
\begin{center}
  
 \includegraphics[width=0.75\textwidth]{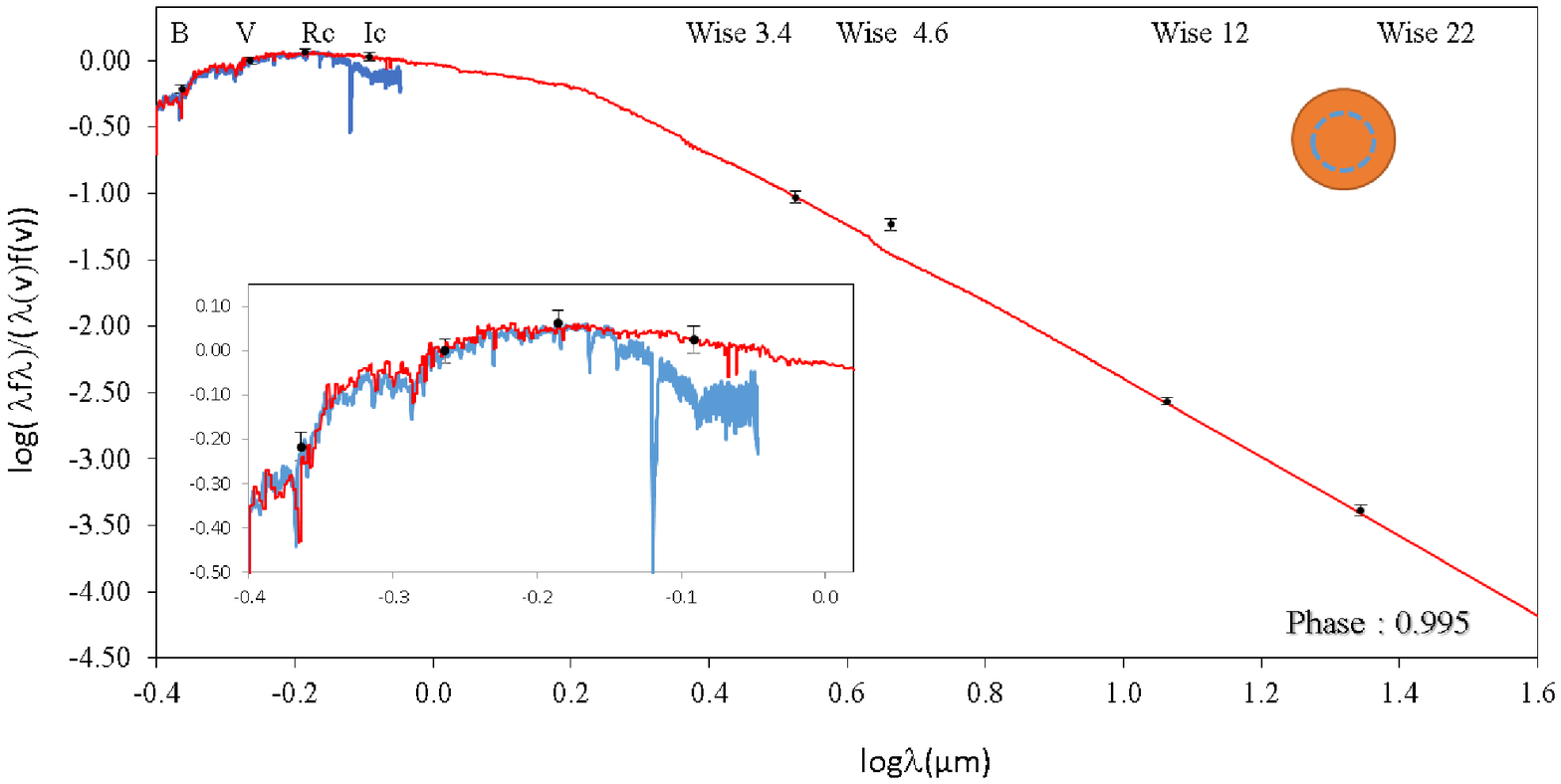}\\
 \includegraphics[width=0.75\textwidth]{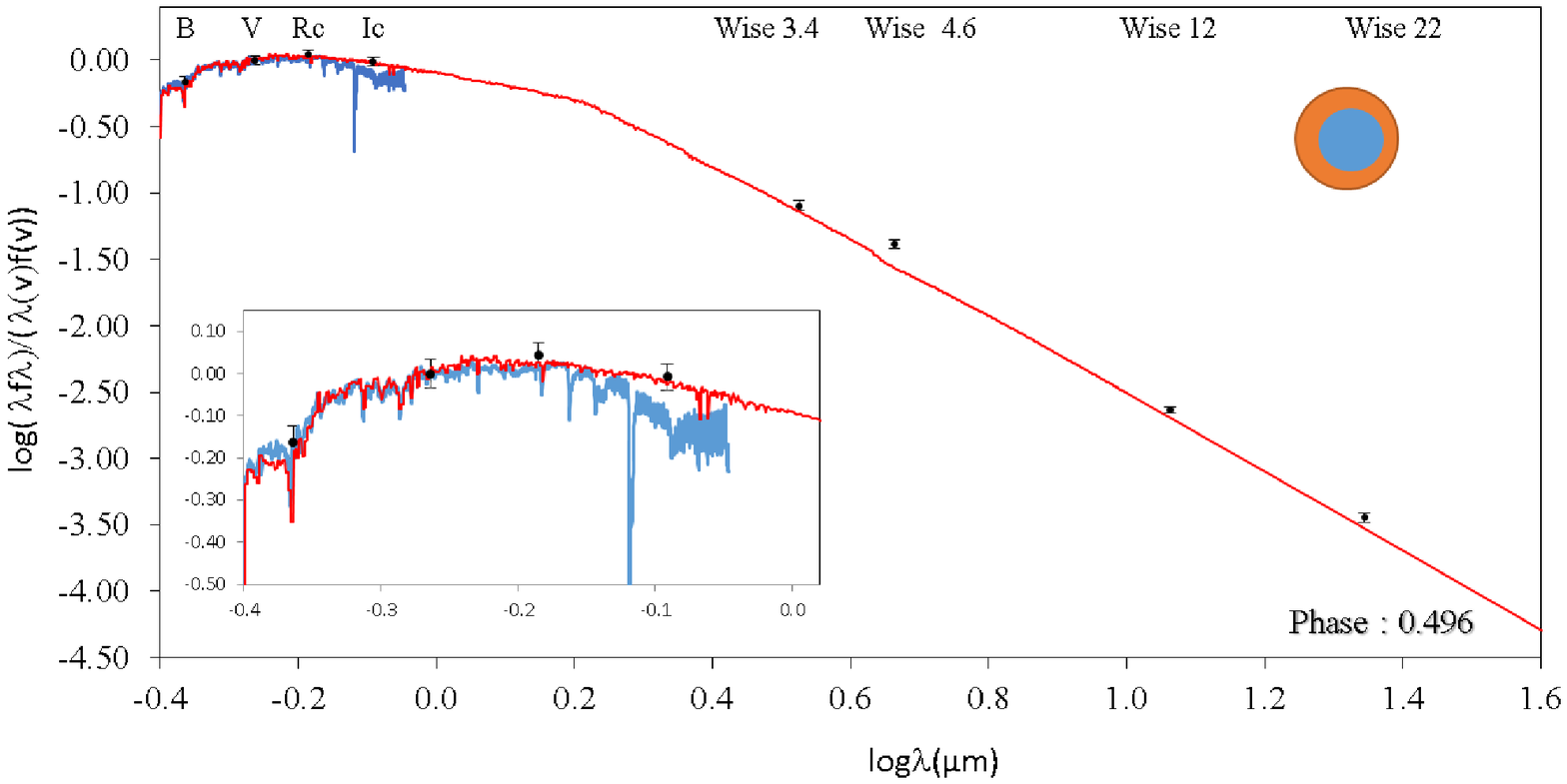}\\
 \includegraphics[width=0.75\textwidth]{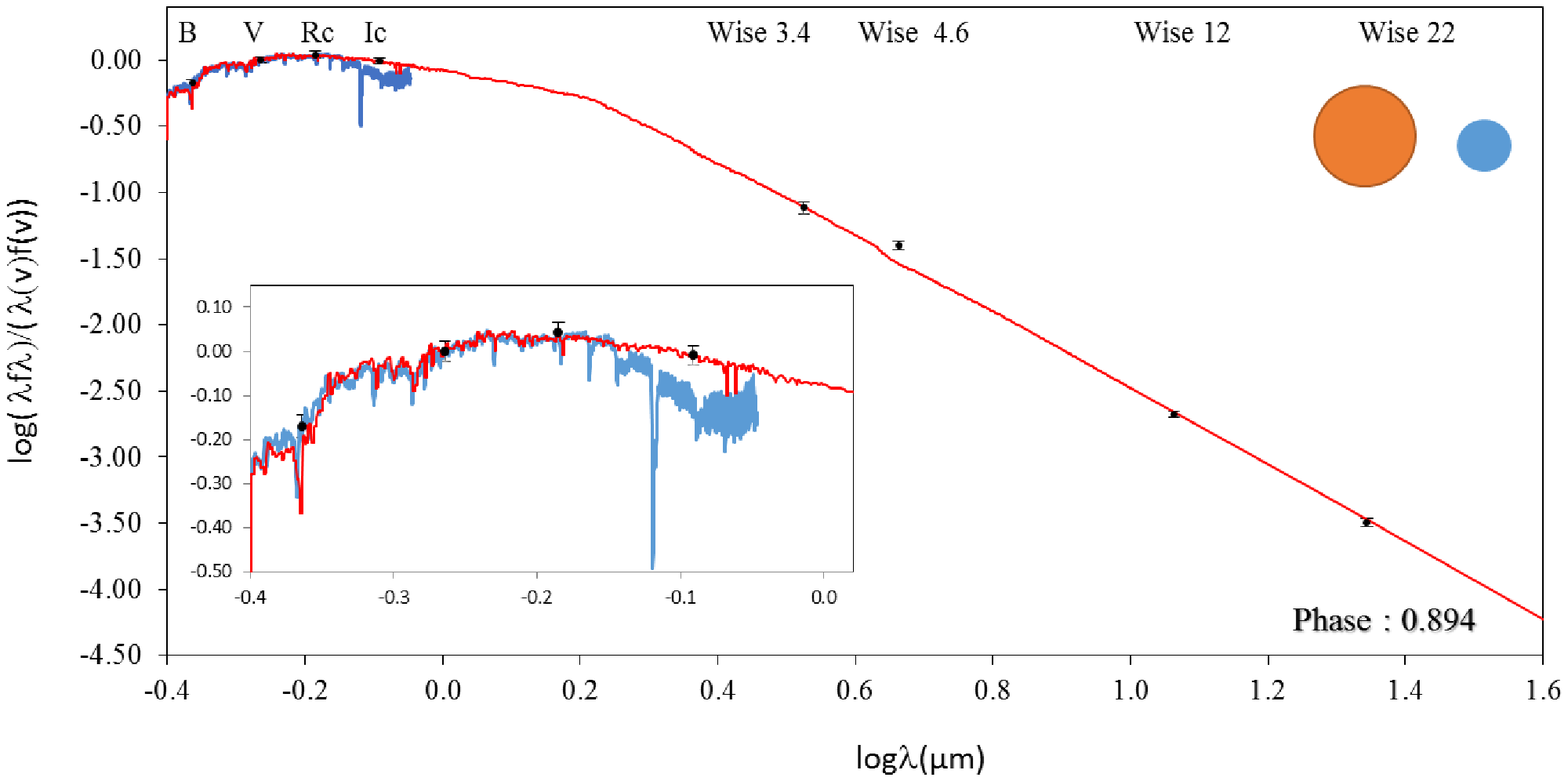}
  
\caption{Spectral energy distribution of AR Lac. Blue line shows spectrum of \texttt{TFOSC}. Red line indicates synthetic spectrum and black dots show photometric data. Blue and orange circles indicate the position of the components.}
\label{fig3}
\end{center}
\end{figure}

As can be seen from Figure~\ref{fig1}, the low resolution \texttt{TFOSC} spectra have the effect of Telluric lines caused by water vapor, oxygen and carbon dioxide molecules in the Earth's atmosphere in the spectral range $6000-9000 \AA$, on the continuum level. This effect is larger in cold stars, especially in the photometric \texttt{"Ic"} band spectral region than in hotter stars.

In order to compare the obtained \texttt{TFOSC} spectra with a synthetic spectrum, the synthetic spectra were constructed by using the appropriate temperature, surface acceleration and metal abundance values, taken from website "http://svo2.cab.intacsic.es/theory/newov2/index.php". Theoretical spectra were obtained by using the ATLAS9 Kurucz \texttt{ODFNEW/NOVER} \citep{Castellietal1997} data in model with zero metal abundance. In constructing a model spectrum of AR Lac binary system, by using the ATLAS9 Kurucz \texttt{ODFNEW/NOVER} model atmosphere, the related parameters were taken as:\\
For the G0 IV component; $T(K)=5750$, $\mathrm{log} \mathit{g}=4.0$, and $Z=0$,\\
For the K0 IV component; $T(K)=5000$, $\mathrm{log} \mathit{g}=3.5$, and $Z=0$.\\
The flux contributions from both components of AR Lac binary system were computed by using the physical parameters (masses; $M_{1,2}$, radii; $R_{1,2}$, Planck functions; $B_{1,2}$, fractional projected area depending on orbital phase; $A_{1,2}(\phi)$, orbital inclination; \textit{i}) as done by \citet{Senavcietal2018} for SV Cam.

In Figure~\ref{fig2}, the spectral energy distribution \texttt{(SED)} modeled for 0.0P orbital phase of AR Lac (i.e. the stellar configuration in which the hotter component of the system is totally eclipsed by cooler component)  was compared with \texttt{SED} modeled for HD 56168, reference star, of the same spectral type with the cooler component of AR Lac. In this Figure~\ref{fig2}, the \texttt{SED} of a \textit{Black Body} radiation with the temperature of the same temperature of the cooler component of AR Lac \texttt{(K0 IV, 5100 K)} was also included. As shown in Figure~\ref{fig2}, the flux values of \texttt{SED} obtained from photometric \texttt{CCD} data are well compatible with the spectral model results. However, this compatibility appears to be seen in the \texttt{B, V, Rc} bands for \texttt{TFOSC} spectral data, while there is a significant decrease in the continuum level of \texttt{TFOSC} spectrum due to Telluric line absorption effect in the \texttt{Ic} band spectral range (see Figure~\ref{fig3}).

\begin{figure}
\vspace{0.0cm}
\begin{center}
  
     \includegraphics[width=0.75\textwidth]{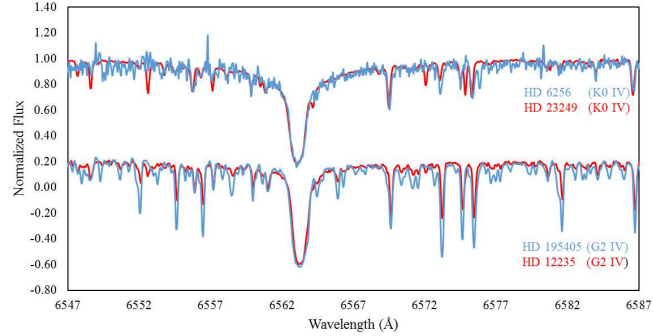}

\caption{Comparison of \texttt{ELODIE} (in red colours) and \texttt{TUG} (in blue colours) H$\alpha$ spectra of reference stars HD 232249 and HR 6256 (at upper)  and HD 12235 and HD195405 (at bottom). In the chart below, the spectra are drawn by subtracting 0.8 from the flux of 1.0.}
\label{fig4}
\end{center}
\end{figure}

\begin{figure}
\vspace{0.0cm}
\begin{center}
  
     \includegraphics[width=0.75\textwidth]{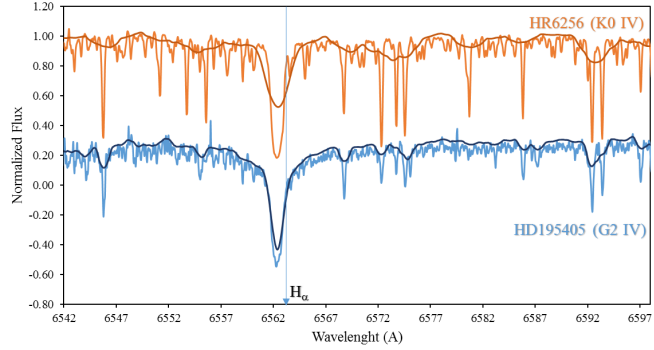}

\caption{Normalized H$\alpha$ spectra of reference stars HD 195405 and HR6256. Straight lines are 30 kms$^{-1}$ for HD 195405 \texttt{(G2IV)} and 70 kms$^{-1}$ for HR6256 \texttt{(K0IV)} refers to the synthetic spectra obtained for rotation speed. In the chart below, the spectrum of HD 195405 was drawn by subtracting 0.7 from the flux of 1.0.}
\label{fig5}
\end{center}
\end{figure}

% ------ Table 4 ------
\begin{table}
\footnotesize
  %\centering 
  \setlength{\tabcolsep}{0.5\tabcolsep} \tablecols{7}
  \caption{Absolute and orbital parameters of AR Lac system}
  \label{tab4}
  \begin{tabular}{lllllll}
  \hline
\textit{Parameters}& \textit{Primary Component}	&\textit{Secondary Component} 	&\textit{Reference}\\
\hline

Spectral Type & \texttt{G2IV}& \texttt{K0IV} &\citet{Frascaetal2000}\\
H$\alpha$ Amplitude of Radial& &	&	 & \\
Velocity Curve (kms$^{-1}$) & $119.43 \pm 0.49$ & $106.73 \pm 0.29$ & \citet{Frascaetal2000}\\ 
M($M_{\odot}$) & $1.17 \pm 0.035$&$1.21 \pm 0.077$ & \citet{Sivieroetal2006}\\
R($R_{\odot}$)	& $1.51 \pm 0.005$ &$2.61 \pm 0.009$ &\citet{Sivieroetal2006}\\
$\mathrm{log} \mathit{g}$ (cms$^{-2}$) & $4.15 \pm 0.021$ & $3.69 \pm 0.035$ & \citet{Sivieroetal2006}\\
T (K) &	$5826 \pm 5$ &$5100 \pm 100$ &\citet{Sivieroetal2006}\\
\textit{v sin i} (kms$^{-1}$) &	46 & 73	&\citet{Frascaetal2001}\\
Orbit Inclination Angle & &	& & \\
\textit{i} (degree)	& 90 & &\citet{Sivieroetal2006}\\
Orbital eccentricity &	&	&	& \\
(e)	& 0	& &\citet{Sivieroetal2006}\\
Space Velocity of the &	& &	& \\
System's Center of Mass & &	& & \\
(kms$^{-1}$) & $-34.54 \pm 0.50$ & &\citet{Frascaetal2000}\\
\hline
  \end{tabular}
   \end{table} 

\subsection{High Dispersion Spectra}
\label{sec:hds}

In order to determine the activity level of a star depending on spectral H$\alpha$ profile, the photospheric effect must be removed from the H$\alpha$ line profile. In the case of binary system, the flux contributions to photospheric H$\alpha$ absorption from both components were also taken into consideration in this method of determination of the activity level. This method, which eliminates the photospheric absorption effect from H$\alpha$ line profile, is called \textit{"Spectral Extraction Method"}. In this study, we tried to analyse the behaviour of H$\alpha$ excess emission in chromospherically active binary system AR Lac, by using this \textit{Spectral Extraction Method}.  For this purpose, high resolution spectra of AR Lac, at different orbital phases, in the wavelength range of $6500-6700 \AA$ with resolution power $R \sim 40000$ were taken on July 13-14, 2013 by using Coude Echelle Spectrograph equipped with \texttt{RTT150} telescope of \texttt{TUG}. The log of high resolution spectral observations are given in Tables~\ref{tab2} and \ref{tab3}.     

% ------ Table 5 ------
\begin{table}
\footnotesize
  \centering 
  \setlength{\tabcolsep}{0.5\tabcolsep} \tablecols{5}
  \caption{Radial velocities from our residual H$\alpha$ profiles of AR Lac active binary system.}
  \label{tab5}
  \begin{tabular}{lllll}
  \hline
  Phase & \multicolumn{2}{c}{\texttt{G2 IV}} & \multicolumn{2}{c}{\texttt{K0 IV}} \\
     & Type & RV (kms$^{-1}$) & Type & RV (kms$^{-1}$) \\
  \hline
  0.013 & - & - & Emission & -28.8 \\
  0.028 & Emission & -47.5 & Emission & -4.6 \\
  0.048 & Emission & -66.7 & Emission & -7.3 \\
  0.065 & Emission & -81.3 & Absorption & -16.9 \\
  0.081 & Emission & -95.5 & Absorption & -47.5 \\
  0.098 & Emission & -116.9 & Absorption & -59.8 \\
  0.514 & Emission & -26.0 & - & - \\
  0.565 & Emission & 23.8 & Emission & -93.2 \\
  0.581 & Emission & 38.4 & Emission & -100.0 \\
  0.598 & Emission & 47.5 & Emission & -109.6 \\
 \hline
  \end{tabular}
   \end{table}

% ------ Table 6 ------
\begin{table}
\footnotesize
  \centering
  \setlength{\tabcolsep}{0.5\tabcolsep} \tablecols{6} 
  \caption{The flux ratios of the components of AR Lac active binary system, in all photometric bands used in this study.}
  \label{tab6}
  \begin{tabular}{lllccc}
  \hline
Band & $\lambda_\textit{$(pivot)$}$(\AA)& Bandwidth (\AA) & Depth of Min. I	(Err)& Depth of Min. II (Err)& $F_{K0IV}$/$F_{G2IV}$ (Err)\\
\hline
\texttt{B} & 4326 & 1816 & 0.491 (0.013) & 0.281 (0.014) & 1.751 (0.057)\\
\texttt{V} & 5445 & 1129 & 0.429 (0.011) & 0.275 (0.012) & 1.562 (0.051)\\
\texttt{R$_{c}$} & 6529 & 1877 & 0.433 (0.012) & 0.311 (0.013) & 1.395 (0.050)\\ 
\texttt{I$_{c}$} & 8104 & 1604 & 0.399 (0.011) & 0.287 (0.011) & 1.388 (0.046)\\
\texttt{W1} & 33526 & 6625.6 & 0.347 (0.033) & 0.291 (0.046) & 1.195 (0.184)\\
\texttt{W2}	& 46028 & 10423 & 0.335 (0.032) & 0.409 (0.023) & 0.821 (0.109)\\
\texttt{W3} & 115608 & 55069 & 0.318 (0.008) & 0.276 (0.007) & 1.151 (0.036)\\
\texttt{W4} & 220883 & 41013 & 0.356 (0.013) & 0.255 (0.011)	& 1.399 (0.059)\\
\hline
  \end{tabular}
   \end{table}
   
\begin{figure}
\vspace{0.0cm}
\begin{center}
  
     \includegraphics[width=0.75\textwidth]{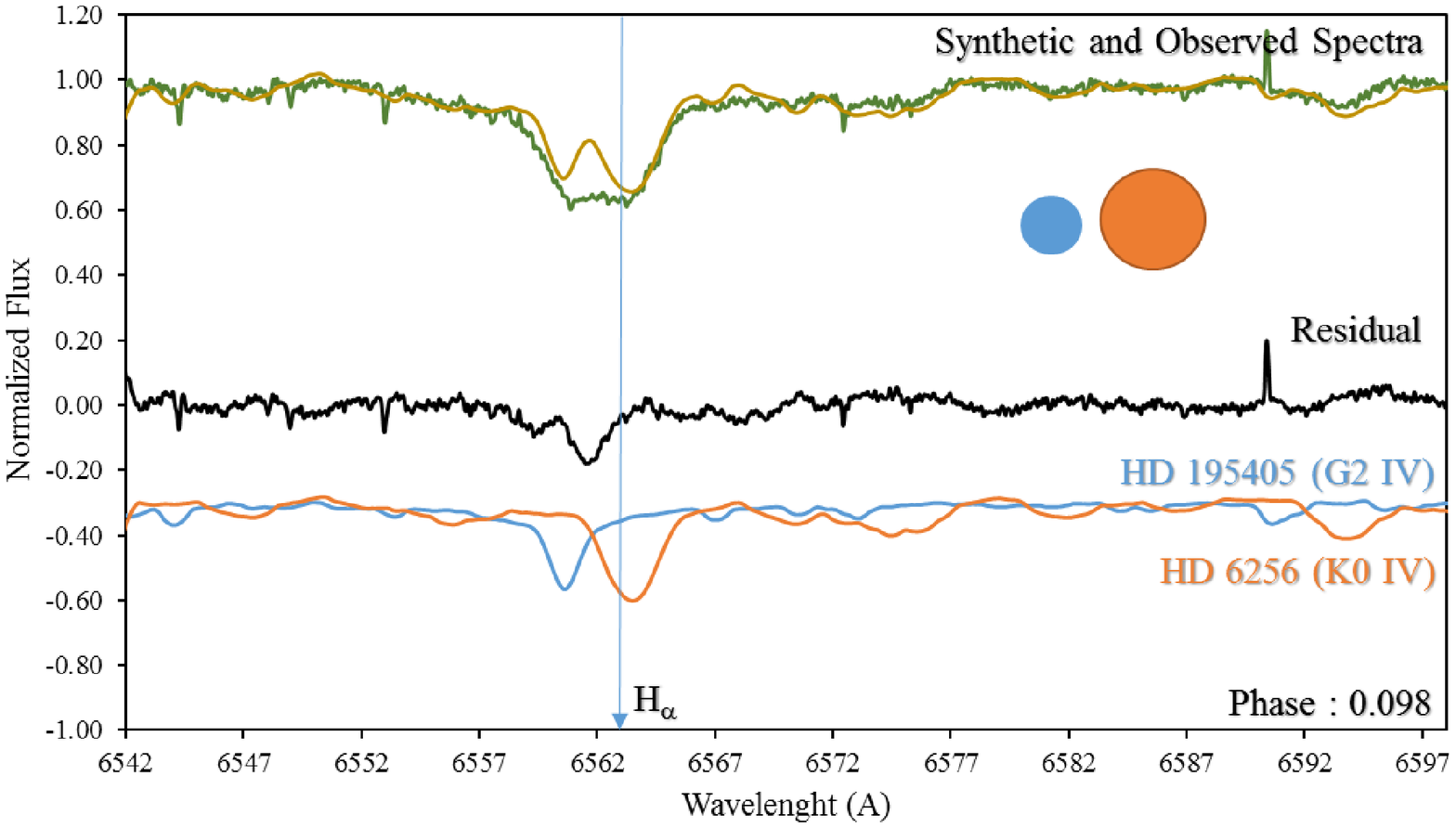}\\
     \includegraphics[width=0.75\textwidth]{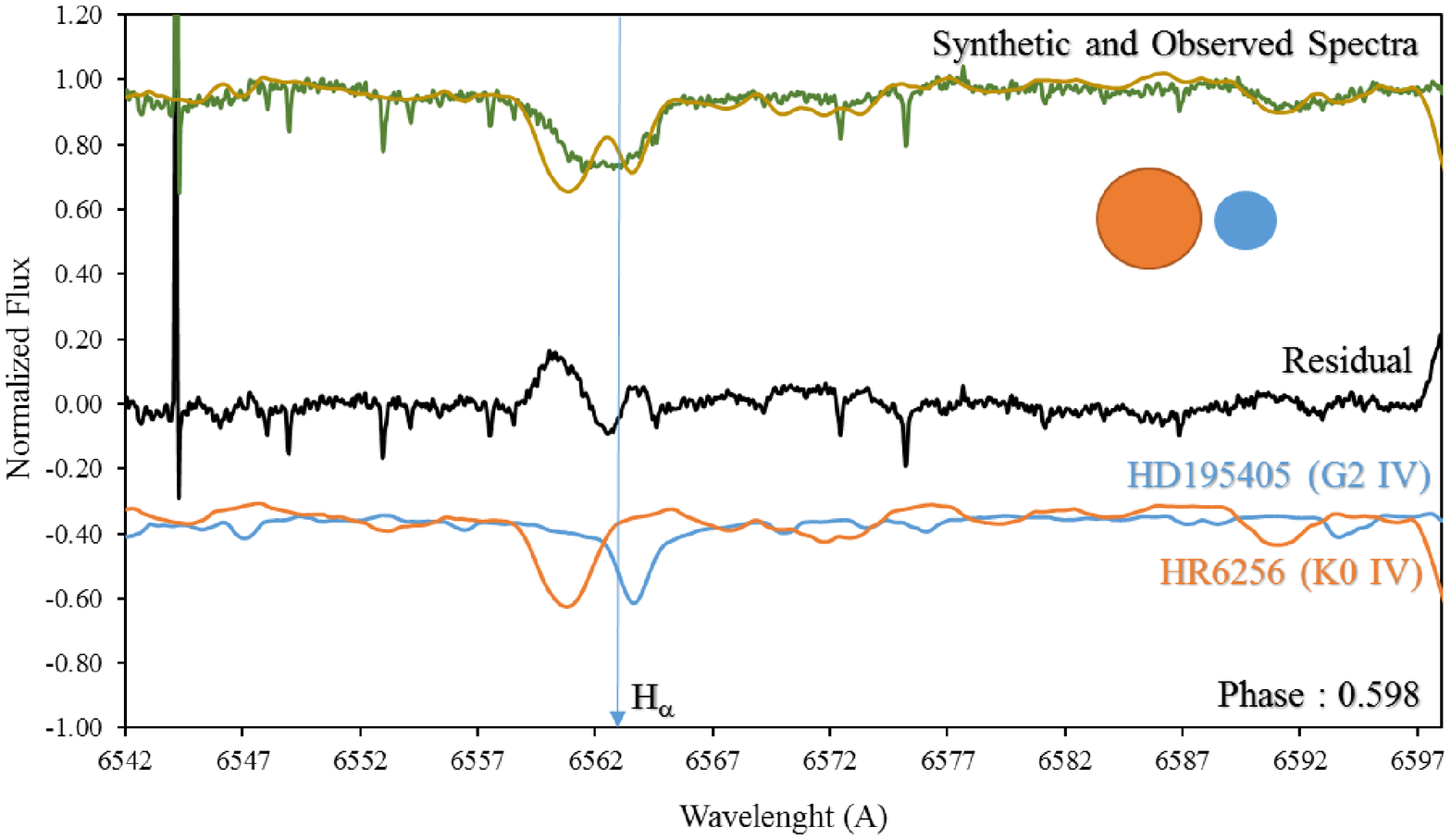}     

\caption{The model spectral solution of high resolution H$\alpha$ of AR Lac. The green line represents the observed spectrum; yellow line shows the model synthetic spectrum; black line shows the residual spectrum; blue and orange lines represent the spectra of reference \texttt{G2 IV} and \texttt{K0 IV} stars, respectively.}
\label{fig6}
\end{center}
\end{figure}

The chromospheric contribution to H$\alpha$ line profile has been determined using the spectral subtraction technique. This technique was applied to high resolution spectra of AR Lac binary system with the following steps:
\begin{itemize}
\item[i)]
The high resolution spectra of reference stars (HD 195405 and HR 6256) taken at \texttt{TUG} were compared with the high resolution \texttt{ELODIE} spectra\footnote{obtained from http://atlas.obs-hp.fr/elodie/index.html} of HD 12235 \texttt{(G2 V)} and HD 23249 \texttt{(K0 IV)} stars \citep[which were selected from the catalogue of][]{Montesetal1997}. This comparison showed that the H$\alpha$ profiles of \texttt{TUG} and \texttt{ELODIE} spectra were compatible (see Figure~\ref{fig4}).

\item[ii)]
From the catalogue of \citet{Montesetal1997}, the rotational equatorial velocities were obtained as < 15 kms$^{-1}$ and 2 kms$^{-1}$ for HD 12235 \texttt{(G2 IV)} and HD 23249 \texttt{(K0 IV)}, respectively. Based on our compatible results, in accordance with their spectral type, we attributed these velocities to our reference stars (HD 195405 and HR 6256).

\item[iii)]
Rotational velocities for the components of AR Lac binary system(\texttt{G2 IV} + \texttt{K0 IV}), were found by \citet{Frascaetal2001} as 46 kms$^{-1}$ and 73 kms$^{-1}$, respectively (see Table~\ref{tab4}). Depending on the orbital phase, the observed H$\alpha$ line profiles of double-lined spectroscopic and active binary system, AR Lac, are shown in the left panel of Figure~\ref{fig6}. Taking into account the rotational velocities of \citet{Frascaetal2001}, given above, the following rotational velocities were applied to the observed spectra of reference stars in modelling the synthetic spectrum of AR Lac binary system by using \texttt{STARMOD} program \citep[see][]{Barden1985, Montesetal2000}:\\
$\textit{V(rot)} \sim 30~kms^{-1}~(=46-15)$, for HD 195405 \texttt{(G2 IV)}\\
$\textit{V(rot)} \sim 70~kms^{-1}~(=73-2)$, for HR 6256 \texttt{(K0 IV)}\\
The input parameters for the \texttt{STARMOD} program were derived by using the parameters of the AR Lac as given in Table~\ref{tab4}.
The obtained synthetic spectra are shown in Figures~\ref{fig5} and \ref{fig6} for reference stars and AR Lac, respectively.
\end{itemize}

\begin{figure}
\vspace{0.0cm}
\begin{center}
  
     \includegraphics[width=1.0\textwidth]{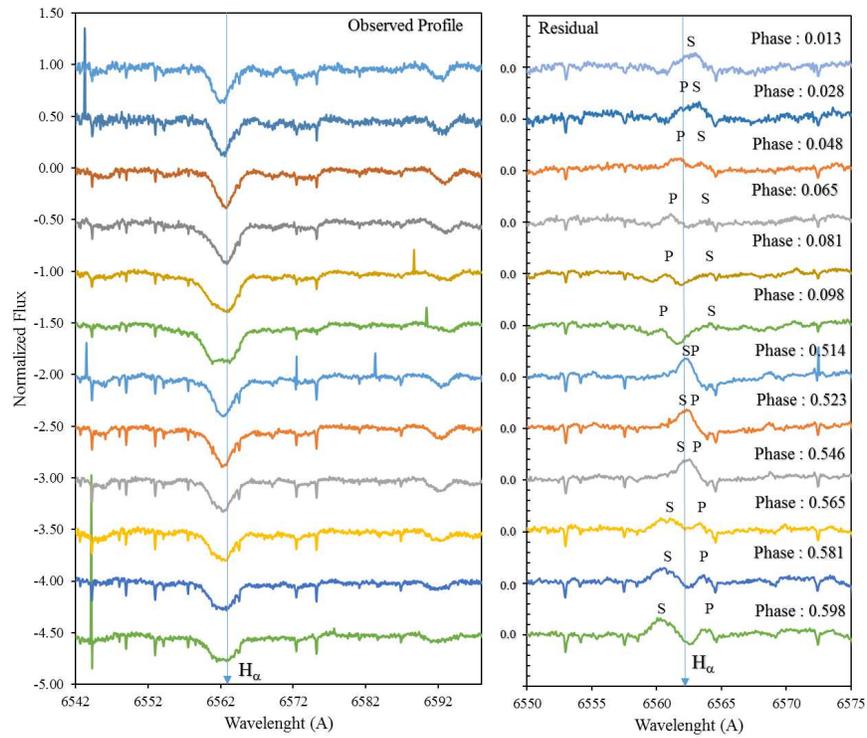}

\caption{High resolution normalized H$\alpha$ line profiles of AR Lac binary system (left panel, consecutive profiles are drawn with a difference of 0.5 in normalized flux values). And, the corresponding residual H$\alpha$ line profiles of AR Lac binary system (right panel; P and S, indicate the positions of the primary/hotter and secondary/cooler components, respectively).}
\label{fig7}
\end{center}
\end{figure}

Since most of the high resolution spectra were taken during minima times within the scope of this study, H$\alpha$ profiles of the components of AR Lac binary system were too close to each other (see Figure~\ref{fig7}). Therefore, the H$\alpha$ profiles of the components could not be separated from each other. In addition, the number of spectral H$\alpha$ profiles of the components that can be distinguished is very few, so the equivalent width measurements for these H$\alpha$ profiles were not made.

\subsection{Photometric CCD and Infrared Observations}
\label{sec:phwo}

In order to evaluate the spectral results, within the scope of extended matter research, of this study together with the results of photometric \texttt{CCD} and Infrared observations, we constructed the \texttt{B, V, R$_{c}$, I$_{c}$}, \texttt{W1, W2, W3} and \texttt{W4} \texttt{(WISE)}\footnote{http://irsa.ipac.caltech.edu/Missions/wise.html} \citep[see][]{Wrightetal2010} band light curves and some characteristics of AR Lac active binary system based on related observational data of \citet{KandE2020}. These \texttt{BVR$_{c}$I$_{c}$} observational data were obtained during the period 2012-2013, while \textit{WISE} data were obtained in 2010. Light curves obtained in normalized luminosity for each band are shown in Figures~\ref{fig9} and \ref{fig10}. And, the results of colour excess \texttt{(CE)} measurements for different photometric bands are also shown in Figure~\ref{fig11}.

By taking the advantage of geometric configuration of the components of an active and total eclipsing binary system, the presence of an excess radiation of active component can easily be detected during the primary minimum at which the occultation of hotter component by cooler and active one \citep[e.g.][]{HallRams1994, KandE2020}. This configuration during primary minimum of a total eclipsing and active binary system could also give an advantage to search for the interaction between the activity/spots phenomena and the extended/circumstellar matter which may exist in the system. Therefore, it is important and useful to examine the structural changes of photospheric/chromospheric spectral line profiles and to reveal the structure in spectral energy distribution \texttt{(SED)}, during primary minimum. In this context, active and eclipsing binaries are important in extended/circumstellar matter's studies.

As can be seen from Figures~\ref{fig9} and \ref{fig10}, in the light curves of AR Lac, the depths of primary and secondary minima vary depending on colour (i.e the photometric band). Using the correlation below for the ratio of the minima depth of a light curve \citep[see][]{KandD1978}, in the case of $\textit{sin i}=1$, we can estimate the flux ratio of the component stars as:

\begin{equation}
\frac {F_{K0IV}}{F_{G2IV}}=\frac{\mathrm{Depth~of~Min.~I}}{\mathrm{Depth~of~Min.~II}},
\end{equation} 
 
Using all these light curves, the results of flux ratios are given in Table~\ref{tab6}.

As can be seen from these observational results, the observed flux ratios decrease gradually by about 10\% towards the longer wavelength \texttt{(W1)}, and in the infrared wavelengths of \texttt{WISE} the flux of components are approximately equal except \texttt{W3} and \texttt{W4} bands (see Table~\ref{tab6}). This decrease in flux ratios towards the longer wavelengths and the equality of fluxes in \texttt{W1} and \texttt{W2} bands suggest that possible extended/circumstellar matter may exist in the AR Lac active binary system.

\section{Results}
\label{sec:Results}

The flux values of \texttt{SED} obtained from photometric \texttt{CCD} data are well compatible with the spectral model results of AR Lac binary system (see Figure~\ref{fig2}). From a comparative analysis of the data in Figure~\ref{fig2}, it was found that during primary minimum of AR Lac, there was an excess radiation of $\simeq 0.143\pm0.043$ in the \texttt{W2} band compared to the HD 56168 reference star. It is clear that this excess radiation at primary minimum can be attributed to the cooler component \texttt{(K0 IV)} of AR Lac binary system.

As can be seen from Table~\ref{tab1}, the low resolution \texttt{TFOSC} spectra of AR Lac were obtained on Oct. 12, 2015 and June 5, 2016 (at times corresponding to minima times), and on Sept. 5, 2016 at times corresponding to outside eclipses. The times which correspond to minima and outside eclipses are determined within the scope of this study. The analysis results of our \texttt{TFOSC} observations were found as given in Figure~\ref{fig3}, along with the synthetic spectrum comparison. In addition, visual and infrared photometric results were added to the evaluation of these low resolution spectral results in Figure~\ref{fig3}.

From the patterns given in Figure~\ref{fig3}, it is seen that the results of low resolution spectra of AR Lac, taken in both minima and outside eclipse phases, are compatible with the results of synthetic spectra. In addition, photometric flux measurements are also observed to be compatible except for the \texttt{W2} band ($\lambda = 46028~\AA$, $\Delta\lambda = 10423~\AA$). The photometric flux measurements in \texttt{W2} band clearly show that there is an excess radiation in this band in the AR Lac. Excess radiation rates in $\mathrm{f(W2)/f(V)}$ have been found as follows:\\
At 0.995P the excess radiation rate is about $0.227\pm0.047$,\\
At 0.496P the excess radiation rate is about $0.176\pm0.031$,\\
At 0.894P the excess radiation rate is about $0.136\pm0.028$.

\begin{figure}
\vspace{0.0cm}
\begin{center}
  
     \includegraphics[width=0.75\textwidth]{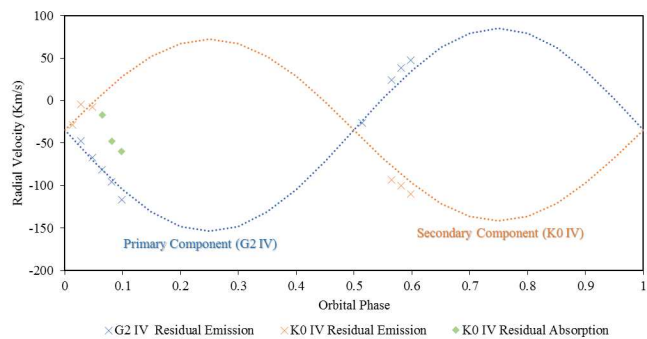}

\caption{A comparison of the radial velocities obtained from our residual H$\alpha$ profiles of AR Lac binary system with the velocity curve by \citet{Frascaetal2000}. The radial velocities of residual H$\alpha$ profiles are denoted with cross signs, and the velocities obtained from the absorption structures appeared in residual H$\alpha$ profiles are also denoted with green squares.}
\label{fig8}
\end{center}
\end{figure}

Based on the model spectral solutions of high resolution H$\alpha$ line profiles (see Figure~\ref{fig6}) of the system, the variation of H$\alpha$ and their residual profiles depending on orbital phases were obtained as given in Figure~\ref{fig7}.

Using our synthetic model spectrum solutions, the radial velocities obtained from residual H$\alpha$ profiles of AR Lac are given in Table~\ref{tab5}. These radial velocity measurements were also compared with the radial velocity curve by \citet{Frascaetal2000}, as given in Figure~\ref{fig8}. The discussion of these results is given in the following Section~\ref{sec:Disc_and_Conc}.

From our analysis of the minimum depth ratios of the light curves of AR Lac (see Table~\ref{tab6} and Figures~\ref{fig9} and \ref{fig10}), obtained in this study, it was seen that the flux values of the components of this active binary system were going to be equal, towards the longer wavelengths. These results are also discussed in Section~\ref{sec:Disc_and_Conc} in the context of existence of an extended/circumstellar matter in the system. 

\section{Discussion and Conclusions}
\label{sec:Disc_and_Conc}

Although recently, an observational evidence of the presence of extended /circumstellar matter has not been found, except in \texttt{W2} band, by \citet{KandE2020} in the AR Lac active binary system based on their \texttt{IR} excess measurement results, this issue has been re-evaluated with the results of the spectral analyses of this study. Our spectral analysis results of the low and high resolution spectra of AR Lac were evaluated within the scope of extended/circumstellar matter researched together with our previous photometric results. A discussion on these results can be summarized as follows:

\begin{itemize}
\item[i)]
During the 0.0P orbital phase of the AR Lac binary system, the hotter component \texttt{(G2 IV)} is totally eclipsed and therefore only the radiation from the cooler component \texttt{(K0 IV)} of the system can be observed. In Figure~\ref{fig2}, the \texttt{SED (Spectral Energy Distributon)} measurement results during 0.0P orbital phase of the AR Lac and the results of the \texttt{HD 56168} ($(B-V)_{\circ}$=$0.90\pm0.01$; as a reference and inactive star) were compared with \texttt{SED} of a black body radiation of an effective temperature of 5100 K. From this comparison, it was seen that the AR Lac has an excess radiation in the \texttt{W2} band. We were unable to compare our \texttt{SED} measurement results with those of \citet{Bussoetal1988} for AR Lac due to their evaluations did not depend on orbital phase and the components of AR Lac were taken into consideration as \texttt{G5 V+G8 IV} (see their Table 5b and Figure 4). However, roughly, our \texttt{SED} results for outside eclipse phases (see the bottom pattern of Figure~\ref{fig3} of this study) appear to be consistent with their results.
\item[ii)]
In addition to our \texttt{SED} evaluations made at the primary minimum for AR Lac binary system in Figure~\ref{fig2}, our \texttt{SED} results, together with our \texttt{TFOSC} spectral data, in the secondary minimum and outside eclipse phases were obtained as given in Figure~\ref{fig3}. From all patterns given in Figure~\ref{fig3}, it clearly appears that AR Lac has a significant excess radiation in the \texttt{W2} band at all orbital phases. The relative value of this excess radiation in the \texttt{W2} band was at highest level at primary minimum and at the lowest level at around outside eclipse phases, with an average value of $\sim 0.18\pm0.03$. This result is also compatible with the results of photometric colour excess \texttt{(CE)} measurements obtained by \citet{KandE2020} for the AR Lac (see Figure~\ref{fig11}). Therefore, these characteristics of excess radiation depending on orbital phase show that the source of excess radiation in the AR Lac is mainly due to the cooler component \texttt{(K0 IV)} of the system. Using the \texttt{Spitzer Space Telescope} data, obtained during the observing period November 2005-January 2007, \citet{Matrangaetal2010} reported that there was no significant warm dust in the AR Lac binary system, but it was found they gave photometric \texttt{IR} measurements that were partially compatible with our findings. Unfortunately, our survey of data sources found no information about which element or molecule is emitting/absorbing a radiation in the region 4.6 microns, and related atomic terms.

\begin{figure}[h!]
\vspace{0.0cm}
\begin{center}
  
     \includegraphics[width=1.0\textwidth]{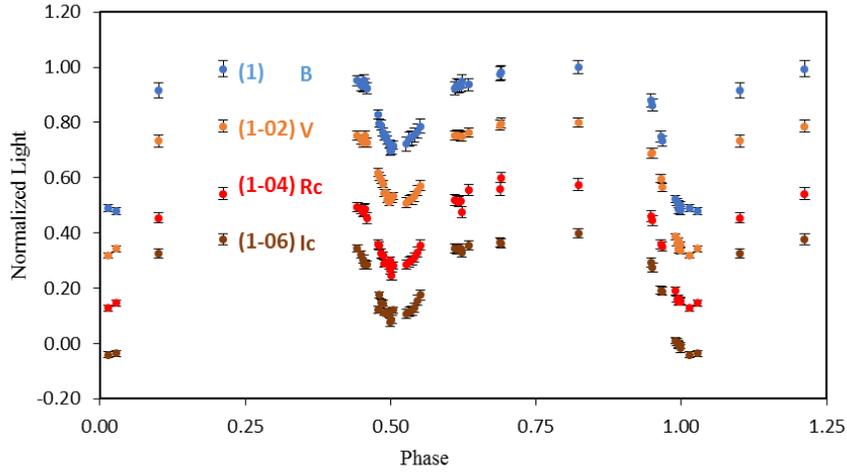}

\caption{Normalized Light curves of AR Lac in \texttt{B, V, R$_{c}$, I$_{c}$} bands.}
\label{fig9}
\end{center}
\end{figure}

\begin{figure}[h!]
\vspace{0.0cm}
\begin{center}
     
     \includegraphics[width=1.0\textwidth]{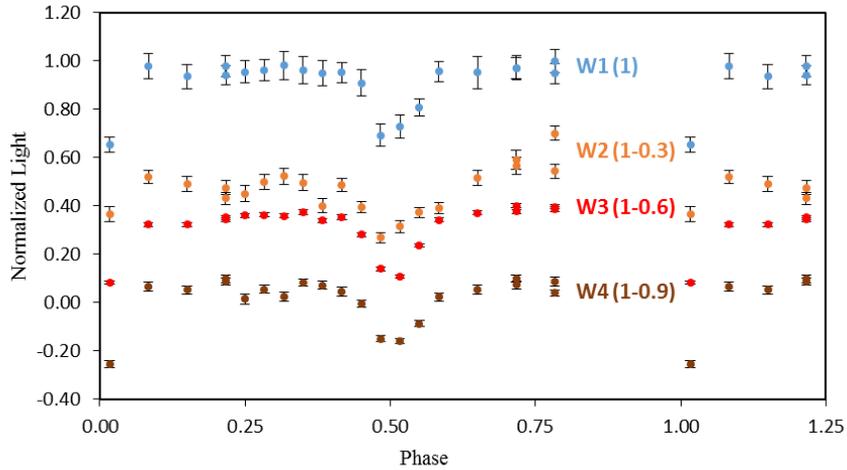}

\caption{Normalized Light curves of AR Lac in \texttt{WISE} bands.}
\label{fig10}
\end{center}
\end{figure}

\item[iii)]
Within the scope of this study, the high resolution H$\alpha$ profile observations of AR Lac were made especially at/near minima times. \textit{The Spectral Extraction Method} was used in modelling spectral H$\alpha$ profiles. The residual H$\alpha$ profiles of AR Lac binary system have been obtained in the form of emission and absorption profiles which are compatible with radial velocities of the components of the system (see Figures~\ref{fig6} and \ref{fig7}). These residual H$\alpha$ profiles are clearly seen as emissions for both components, especially at 0.013P, 0.028P, 0.514P, 0.546P, 0.565P, and 0.581P orbital phases. However, the central depths of H$\alpha$ absorption profiles of AR Lac at 0.048P, 0.065P, 0.081P, and 0.098P orbital phases were less deeper than of the depths of synthetic profiles (see Figure~\ref{fig6}). The absorption feature in the residual H$\alpha$ profiles of AR Lac could be seen at these orbital phases. This excess absorption structure in the residual H$\alpha$ profiles of AR Lac observed at/near primary minimum phases is compatible with the finding results of \citet{Zboriletal2004} and \citet{Frascaetal2000}. These residual H$\alpha$ profiles (see Figure~\ref{fig7}) provide information about the level and the variation of stellar activity in the AR Lac binary system. In addition, the residual H$\alpha$ profiles of the cooler component of AR Lac binary system appear to be wider and intensive than the profiles of the hotter component. That is, from these residual H$\alpha$ profiles, given in Figure~\ref{fig7}, we see that the stellar activity in AR Lac binary system comes mainly from the secondary/cooler component and this cooler component rotates faster than primary/hotter component. These are the characteristics related to the expected results of chromospheric active stars.
\item[iv)]
Using synthetic spectra constructed for high resolution Coude Echelle spectra of AR Lac binary system, the radial velocity measurements of the component stars were made from the residual H$\alpha$ profiles. All these radial velocity measurement results are shown in Figure~\ref{fig8} together with the radial velocity curves drawn based on the radial velocity measurement results obtained by \citet{Frascaetal2000} for the AR Lac binary system. As can be seen from Figure~\ref{fig8}, the differences between these radial velocities \textit{(RVs)} and the photospheric \textit{RVs} give us observational evidence about the \textit{RVs} which correspond to active regions in the chromosphere, which are at higher layers of the stellar atmosphere. \citet{Frascaetal2000}, based on their study on H$\alpha$ spectroscopy of AR Lac, had reported an extra absorption during primary eclipse (at 0.99P)  which extends in blue side producing an asymmetric emission of the cool star with a center position shifted to red by 68 kms$^{-1}$ (i.e. equal to the \textit{v sin i} of the \texttt{K0 IV} component). A similar result of this extra absorption was determined by using the residual H$\alpha$ absorption structure with the \textit{RVs} of about -47.5 kms$^{-1}$ and -59.8 kms$^{-1}$, at 0.081P and 0.098P, respectively (see Table~\ref{tab5} and green squares in Figure~\ref{fig8}). It seems that this extra absorption is most likely due to the prominence on the \texttt{K0 IV} component of AR Lac binary system. In summary, based on this observational evidence, it appears that a prominence-like extended/circumstellar matter most likely exist around the cooler component of AR Lac active binary system. Therefore, it can be suggested that such a low-density extended/circumstellar matter causes a colour excess or a residual emission/absorption in the AR Lac binary system.
\item[v)]
Our evaluations on the flux ratios of the component stars of AR Lac from the minima depth ratios of the light curves of the system (see Table~\ref{tab6} and Figures~\ref{fig9} and \ref{fig10}), show that the fluxes of the component stars are to be equal or almost equal to each other in \texttt{W1, W2}, and \texttt{W3} band light curves. As can be seen from Table~\ref{tab6}, the flux ratio or the component of AR Lac in \texttt{W2} band is smaller than 1. This is because the depth of primary minimum(\texttt{Min. I}) of the light curve is less than of secondary minimum (\texttt{Min. II}), that is, when the hotter component(\texttt{G2 IV}) is behind the cooler component(\texttt{K0 IV}). In other words,in \texttt{W2} band the system is brighter during primary minimum than in its secondary minimum phases. On the other hand, it was found that \texttt{W2} band is additionally sensitive to hot dust\citep[see][]{Cluveretal2014}. Therefore, the fact that the \texttt{W2} flux ratio for the components of AR Lac binary system is smaller than 1, suggests that the extended/circumstellar matter around \texttt{K0 IV} component could likely be heated by the hotter component during the orbital phase 0.0P. In addition, the equality in fluxes, for the remaining bands, could be the result of an extended/circumstellar matter/material that can be detected in the wavelengths of these bands in the AR Lac. The colour excess\texttt{(CE)} measurement results in these bands by \citet{KandE2020} also support this suggestion: \textit{CE} values in these bands were obtained as $\mathrm{CE(V-W1)}\simeq \mathrm{CE(V-W3)} \simeq 0.1$ and $\mathrm{CE(V-W2)} \simeq 0.4$ (see Figure~\ref{fig11}).
\end{itemize} 

\begin{figure}
\vspace{0.0cm}
\begin{center}
  
     \includegraphics[width=0.75\textwidth]{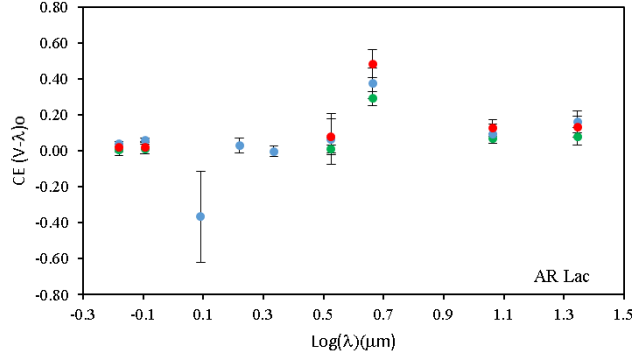}
  
\caption{As a function of wavelength, the colour excesses of AR Lac in all bands during minima and outside eclipses (0.0P are indicated as red colour, 0.5P as blue, and the outside phases as green points). \texttt{CE} values are in magnitudes \citep[from][]{KandE2020}.}
\label{fig11} 
\end{center}
\end{figure} 

The results of radio and polarization observations of AR Lac binary system, published in the period 1977-2007 \citep[see][]{OandS1977, Brownetal1979, Walteretal1983, DandM1984, Walteretal1987, Trigilioetal2001, Koch2007}, are in agreement with the results of this study. Based on all these spectroscopic and photometric findings obtained in this study, it can be concluded that in the AR Lac, there are not only evidences/contributions of excess radiation caused by stellar activity, but also caused by the presence of an extended/circumstellar matter. In other words, the results obtained in this study, together with the radio and polarization observation results of AR Lac binary system, were evaluated as follows: We made an investigation to find an evidence as to whether the effect of the extended/circumstellar matter(thought to exists) or the extended corona of the \texttt{K0 IV} component of the system contributes more to emergence of these observational findings. 

It can be clearly seen that these results are in agreement with the conclusion of this study that an extended/circumstellar matter could, most likely, exist in the AR Lac active binary system.

However, it should also be noted that although the mass-loss rate obtained by using the long-term secular period decreasing due to stellar magnetic activity of AR Lac binary system is very small \citep[see][]{Luetal2012}, we see that the stellar activity of this binary system is sufficient to affect the coronal structure \citep[see][]{Walteretal1983} and the minima depth ratios of the light curves (see Table~\ref{tab6}, Figures~\ref{fig9} and ~\ref{fig10}; which suggest the presence of common envelope such as $\beta$ Lyr or W UMa type binary systems) and the residual H$\alpha$ emissions (see Figure~\ref{fig7}) gave some important observational evidences for extended/circumstellar matter. Therefore, it is useful to have some more sensitive observational studies and continue these researches.

We thank to T\"{U}B\.{I}TAK National Observatory for a partial support in using \texttt{RTT150} telescopes with project numbers 13ARTT150-406 (Coud\'{e}) and 14BRTT150-664 \texttt{(TFOSC)}. We would like to thank Prof. Dr. \.{I}lbeyi A\u{g}abeyo\u{g}lu for checking out English text. And finally, we would like to thank the referee for his/her directions on some points to improve the comments of some results of this study. This research has made use of the Simbad Database operated at CDS, Starsbourg, France and of NASA's Astrophysics Data System Bibliographic Services. This work has also made use of data from European Space Agency \texttt{(ESA)} mission \texttt{Gaia}(https://www.cosmos.esa.int/gaia), produced by the \textit{Gaia} Data Processing and Analysis Consortium (DPAC, https://www.cosmos.esa.int/web/).

\end{document}